\documentclass[letterpaper,english,reprint, aps,prb,showkeys,floatfix]{revtex4-1}
\pdfoutput=1
\usepackage[T1]{fontenc}
\usepackage[latin9]{inputenc}
\usepackage{array}
\usepackage{booktabs}
\usepackage{multirow}
\usepackage{amsmath}
\usepackage{amssymb}
\usepackage{graphicx}
\usepackage{subscript}
\usepackage{color}

\makeatletter

\pdfpageheight\paperheight
\pdfpagewidth\paperwidth

\providecommand{\tabularnewline}{\\}

\makeatother

\usepackage{babel}
\begin{document}
\preprint{APS/123-QED}
\title{Excitonic Magneto-Optical Kerr Effect in 2D  Transition Metal Dichalcogenides Induced by Spin Proximity}
\author{J. C. G. Henriques}
\affiliation{Department and Centre of Physics, and QuantaLab, University of Minho,
Campus of Gualtar, 4710-057, Braga, Portugal}
\author{G. Catarina, A. T. Costa, and J. Fern{\'a}ndez-Rossier}
\affiliation{International Iberian Nanotechnology Laboratory (INL), Av. Mestre
	Jos{\'e} Veiga, 4715-330, Braga, Portugal}
\author{N. M. R. Peres}
\email{ peres@fisica.uminho.pt, nuno.peres@inl.int}
\affiliation{Department and Centre of Physics, and QuantaLab, University of Minho,
	Campus of Gualtar, 4710-057, Braga, Portugal}
\affiliation{International Iberian Nanotechnology Laboratory (INL), Av. Mestre
	Jos{\'e} Veiga, 4715-330, Braga, Portugal}
\date{\today}
\begin{abstract}
In this paper we develop the excitonic theory of Kerr rotation angle in a two-dimensional (2D)
transition metal dichalcogenide at zero magnetic field. The finite Kerr angle is induced  by  the interplay between spin-orbit splitting and proximity exchange coupling due to the presence
of a ferromagnet.   We compare the excitonic effect with the single particle theory approach. We show that the excitonic
properties of the 2D material lead to a dramatic change in the frequency dependence  of the optical response
function. We also find that the excitonic corrections enhance the optical response by  a factor of two in the case of MoS2 in proximity to a Cobalt thin film.   
\end{abstract}
\maketitle

\section{Introduction}
\label{sec:Introduction}

Proximity effects have been known for decades \cite{holm1932messungen},
but their true potential was only  unleashed  
with the rise of two dimensional
(2D) materials \cite{novoselov20162d}. When working with bulk materials,
proximity effects are negligible, since the size scale of the material
is many orders of magnitude superior to the scale along which proximity
effects are noticeable, relegating them to localized phenomena occuring within
few nanometers \cite{buzdin2005proximity}. Working with 2D materials,
however, is a strikingly different scenario. Their low dimensionality
is responsible for the enhancement of proximity effects. After all,
their thickness can be orders of magnitude inferior to the length
scale of those effects. This allows the wave
function of the material causing the proximity effect to totally engulf the 2D system \cite{vzutic2019proximitized},
thus drastically modifying its intrinsic properties. These effects
are responsible for inducing new features in the adjacent regions, such as, turning
a non-magnetic material into a magnetic one, or giving rise to topologically
non-trivial properties where otherwise there were none \cite{buzdin2005proximity,wang2015proximity,fu2008superconducting,vzutic2019proximitized}.

\begin{figure}
	\includegraphics[width=8cm]{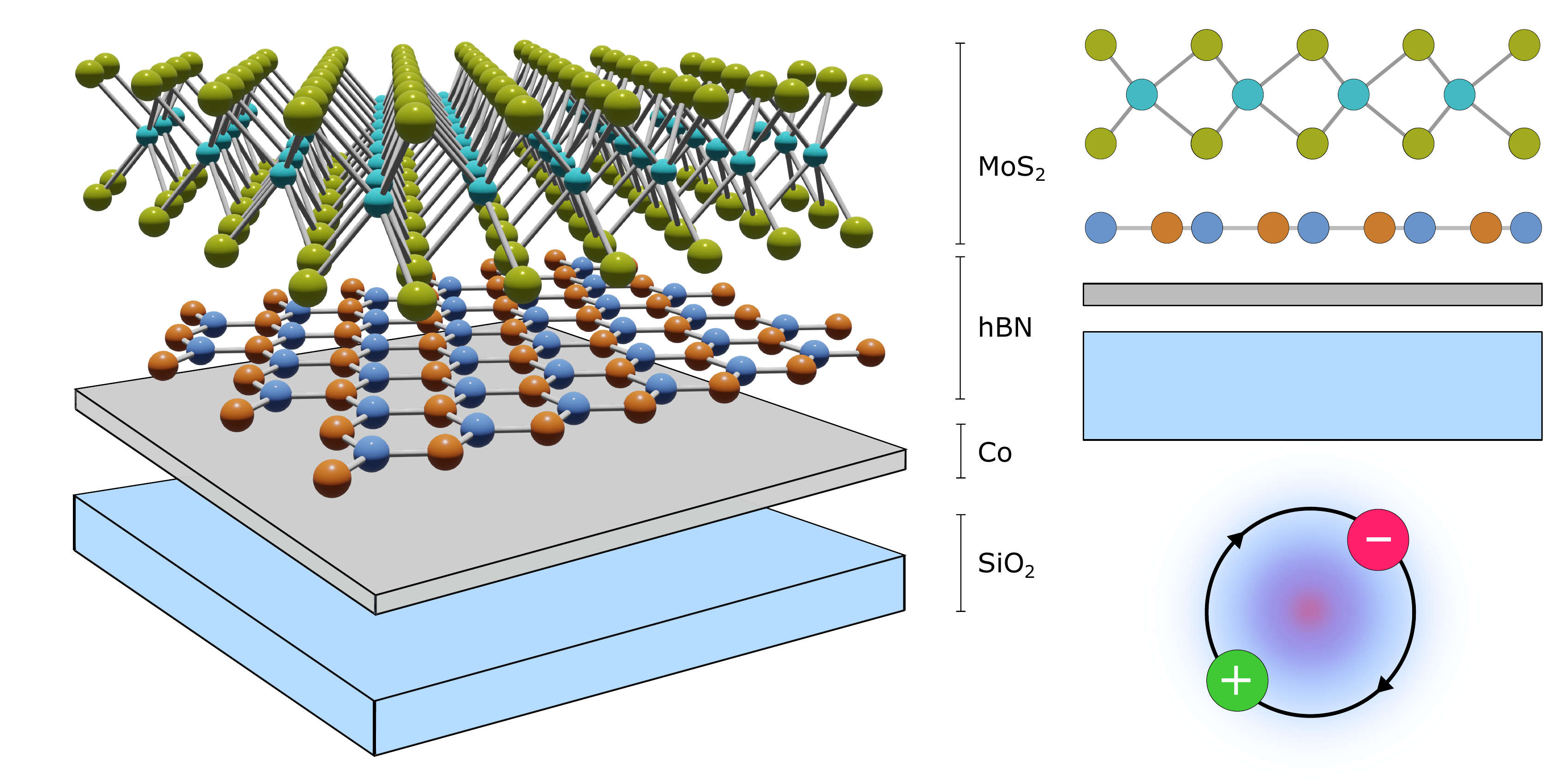}
	\caption{\label{fig:TMD lattice}
		Representation of the system considered in this paper:  a heterostructure composed of MoS$_2$/hBN/Cobalt thin film/quartz; both a three dimensional perspective and a transverse cut view of the heterostructure are shown. 
		The hBN is single layer and acts as a buffer layer to protect MoS$_2$  from direct contact with the metallic Cobalt thin film (3 layers). An artistic view of an exciton formed in MoS$_2$ by the impinging electromagnetic radiation is also depicted. The band structure of this heterostructure was first discussed in Ref. [\onlinecite{Zollner2019}].}
\end{figure}

Among the large variety  of 2D materials, transition metal dichalcogenides
(TMD) are some of the most prominent and studied ones \cite{wang2018colloquium}.
A monolayer TMD is composed by a layer of transition metal atoms,
situated between two layers of chalcogen atoms, forming a trigonal
prism structure. A representation of the real lattice of a generic
TMD is given in Fig. \ref{fig:TMD lattice}. Contrary to graphene,
the existence of different types of atoms in each sub-lattice, leads
to the opening of gaps at the corners of the first Brillouin zone.
According to Ref. [\onlinecite{chaves2017excitonic}] these band gaps are
of the order of $1$ eV. Another aspect in which TMD's differ from
graphene, is the existence of strong spin orbit coupling (SOC), due
to the presence of heavy atoms with $d$ orbitals. To describe the
band structure of these materials, one uses a massive Dirac Hamiltonian,
to which a SOC term must be added. The existence of spin orbit coupling
is also responsible for the coupling of spin and valley, which leads
to valley selective helicity in interband transitions \citep{xiao2012coupled}. A key aspect
to control the valley degrees of freedom is to break the existing
symmetry between the point $K$ and $K'$ at the corners of the first
Brillouin zone. Unfortunately, this proves to be an extremely hard
task when using an external magnetic field, since fields as large as $\sim10$
T are necessary to produce a minute splitting of $\sim1$ meV \cite{macneill2015breaking, stier2016exciton, arora2016valley, zhao2017enhanced, zhong2017van, seyler2018valley}.

It is at this point that a proximitized TMD becomes an entirely new
system with the desired properties, since it has been shown that the
proximity with a magnetic material produces the needed valley splittings magnitudes \cite{niu2015giant}. Previous works have already studied the valley manipulation due to proximity with antiferromagnetic \cite{xu2018large,li2018large} and ferromagnetic \cite{zhang2016large,qi2015giant,zhao2017enhanced} substrates, as well as the effect of proximity with CrI\textsubscript{3}, an ultrathin ferromagnetic semiconductor \cite{Zollner2019b,zhong2017van,seyler2018valley}.

With the ability to control both spin and valley degrees of freedom
by ingeniously choosing an adequate substrate, we can explore magneto-optical
effects, such as the Kerr rotation angle in the absence of magnetic fields. Although these kind of effects
are vastly studied and used in bulk materials \cite{cebollada1994enhanced,zeper1989perpendicular,oppeneer1992ab},
their true potential in 2D materials is yet to be fulfilled. In this
work, we discuss the effect that proximity
induced phenomena have in the band structure and optical conductivity (both longitudinal and Hall)
of a TMD. We show that the Hall conductivity becomes
non-zero when the valley symmetry is broken due to proximity with
a magnetic thin film. The finite Hall conductivity is the key feature
to obtain a non-zero Kerr rotation angle.

This work is organized as follows:
we start in Sec. \ref{sec:Model-Hamiltonian} by introducing the model Hamiltonian
 that will be used throughout
the text, and apply it to the case of MoS\textsubscript{2} on a heterostructure composed of MoS\textsubscript{2}/hBN single layer/Cobalt thin film/quartz. Afterwards
in Sec. \ref{sec:Optical-Conductivity}
 we compute the absorbed power of a 2D material from Fermi golden rule,
for both linear and circular polarized light, and establish its relation
with different entries of the conductivity tensor. Finally 
in Sec. \ref{sec:Excitonic-Effects-in}
we apply this formalism to
the study of the Hall conductivity, and later to the Kerr angle, in the
presence of excitonic effects. An appendix gives the transformation of the Bethe-Salpeter equation to
real space.

\section{Model Hamiltonian and Band Structure\label{sec:Model-Hamiltonian}}

In this section, we introduce the model Hamiltonian that will be used
throughout the text, composed by the standard  Dirac Hamiltonian with finite mass to which
a spin-orbit contribution and an exchange term is
added (numerical value of the latter has been determined from {\it ab-initio} calculations). We will be specific and consider the case of monolayer MoS\textsubscript{2}
heterostructure (see Fig. \ref{fig:TMD lattice}) composed of MoS\textsubscript{2}/hBN single layer/Cobalt thin film/quartz, a system for which a giant magnetic exchange was found \cite{Zollner2019}. Following that recent work \cite{Zollner2019}, the Cobalt thin film is composed of the three layers. The hBN single layer is used as buffer layer and the Cobalt provides the proximity-induced exchange. Since both the Cobalt and the hBN are extremely thin, we assume that most of the screening of the electric field between the electron and the hole is 
provided by the quartz (or otherwise) substrate.

\subsection{Model}
\label{subsec:Model}

To describe our system we adopt a low-energy effective Hamiltonian
with the structure $H=H_{0}+H_{SOC}+H_{ex}$, where $H_{0}$ is the
usual Dirac Hamiltonian, $H_{SOC}$ describes the spin orbit coupling,
and $H_{ex}$ characterizes the exchange splitting due to magnetic
proximity effects. When written explicitly, and in agreement with
Refs. \cite{da2017nonreciprocal,niu2015giant,scharf2017magnetic,xiao2012coupled},
the total Hamiltonian is given by
\begin{align}
H & =v_{F}\hbar(\tau k_{x}\sigma_{x}+k_{y}\sigma_{y})+\frac{m}{2}\sigma_{z}\nonumber \\
& +\tau s_{z}\Bigg(\lambda_{c}\frac{1+\sigma_{z}}{2}+\lambda_{v}\frac{1-\sigma_{z}}{2}\Bigg)\nonumber \\
& -s_{z}\Bigg(B_{c}\frac{1+\sigma_{z}}{2}+B_{v}\frac{1-\sigma_{z}}{2}\Bigg),
\end{align}
where $v_{F}$ is the Fermi velocity; $\hbar$ is the reduced Planck's
constant; $\tau=\pm1$ is the valley index referring to the $K$ and
$K'$ valleys, respectively; $s_{z}=\pm1$ is the spin index labeling
spin up and spin down, respectively; $\sigma_{x},$ $\sigma_{y}$,
$\sigma_{z}$ are $\text{2\ensuremath{\times2}}$ Pauli matrices;
$k_{x}$ and $k_{y}$ are the $x$ and $y$ components of the wave
vector $\mathbf{k}$; $m$ is the band gap when no other contributions
are considered; $\lambda_{c}$ and $\lambda_{v}$ characterize the
spin orbit coupling splitting in the conduction and valence band;
$B_{c}$ and $B_{v}$ describe the effective exchange splitting for
the conduction and valence band respectively, induced by proximity (these parameters can be determined using {\it ab-initio} methods, as noted before).

Solving  the eigenproblem $H|u_{\alpha}^{\tau,s_{z}}\rangle=E_{\alpha}|u_{\alpha}^{\tau,s_{z}}\rangle$,
with $\alpha=\{c,v\}$ (or $\alpha=\pm1,$ respectively, depending
on the context), one easily obtains
\begin{equation}
|u_{c}^{\tau,s_{z}}\rangle=\left(\begin{array}{c}
\frac{\mathcal{M}_{c}^{\tau,s_{z}}}{\sqrt{(\mathcal{M}_{c}^{\tau,s_{z}})^{2}+v_{F}^{2}\hbar^{2}k^{2}}}\\
\frac{\tau v_{F}\hbar ke^{\tau i\theta}}{\sqrt{(\mathcal{M}_{c}^{\tau,s_{z}})^{2}+v_{F}^{2}\hbar^{2}k^{2}}}
\end{array}\right),\label{eq:Conduction Spinor}
\end{equation}
and
\begin{equation}
|u_{v}^{\tau,s_{z}}\rangle=\left(\begin{array}{c}
\frac{-\tau v_{F}\hbar ke^{-\tau i\theta}}{\sqrt{(\mathcal{M}_{v}^{\tau,s_{z}})^{2}+v_{F}^{2}\hbar^{2}k^{2}}}\\
\frac{\mathcal{M}_{v}^{\tau,s_{z}}}{\sqrt{(\mathcal{M}_{v}^{\tau,s_{z}})^{2}+v_{F}^{2}\hbar^{2}k^{2}}}
\end{array}\right),\label{eq:Valence Spinor}
\end{equation}
with $\tan \theta=k_y/k_x$, and $\mathcal{M}_{c}^{\tau,s_{z}}=m/2+B_{v}s_{z}-\lambda_{v}s_{z}\tau+E_{c}$,
and $\mathcal{M}_{v}^{\tau,s_{z}}=m/2-B_{c}s_{z}+\lambda_{c}s_{z}\tau-E_{v}$,
considering
\begin{equation}
E_{\alpha}=s_{z}\Big((\lambda_{c}+\lambda_{v})\tau-(B_{c}+B_{v})\Big)+\alpha\sqrt{\bigg(\frac{\zeta_{\tau,s_{z}}}{2}\bigg)^{2}+v_{F}^{2}\hbar^{2}k^{2}},\label{eq:Energy Spectrum}
\end{equation}
where $\zeta_{\tau,s_{z}}=m+s_{z}(B_{v}-B_{c}+\lambda_{c}\tau-\lambda_{v}\tau)$
is the gap between the valence and conduction band for a specific
choice of $\tau$ and $s_{z}$.

The total electronic Hamiltonian is the sum of $H$ with the Rytova-Keldysh
potential, defined as \cite{keldysh1979coulomb,rytova1967}
\begin{equation}
V(r)=\frac{e^{2}}{4\pi\epsilon_{0}}\frac{\pi}{2}\frac{1}{r_{0}}\Bigg[\mathbf{H}_{0}\bigg(\frac{\kappa r}{r_{0}}\bigg)-Y_{0}\bigg(\frac{\kappa r}{r_{0}}\bigg)\Bigg],\label{eq:Rytova-Keldysh Potential}
\end{equation}
where $r_{0}\sim d\epsilon/2$, with $d$ and $\epsilon$ the thickness
and dielectric function of the 2D material, respectively (microscopically, $r_0$ relates to the polarizability of the 2D system); $\kappa$
is the mean dielectric function of the media surrounding the 2D material; $\epsilon_0$ is the vacuum permittivity; $e$ is the elementary charge; 
$\mathbf{H}_{0}(x)$ is the Struve function, and $Y_{0}(x)$ is the Bessel
function of the second kind. This potential is the solution of the
Poisson equation for a thin film embedded in a medium.

\subsection{Band structure of the  heterostructure of MoS\protect\textsubscript{2} }
\label{subsec:MoTe-on-EuO}

\begin{figure}
		\includegraphics[width=8cm]{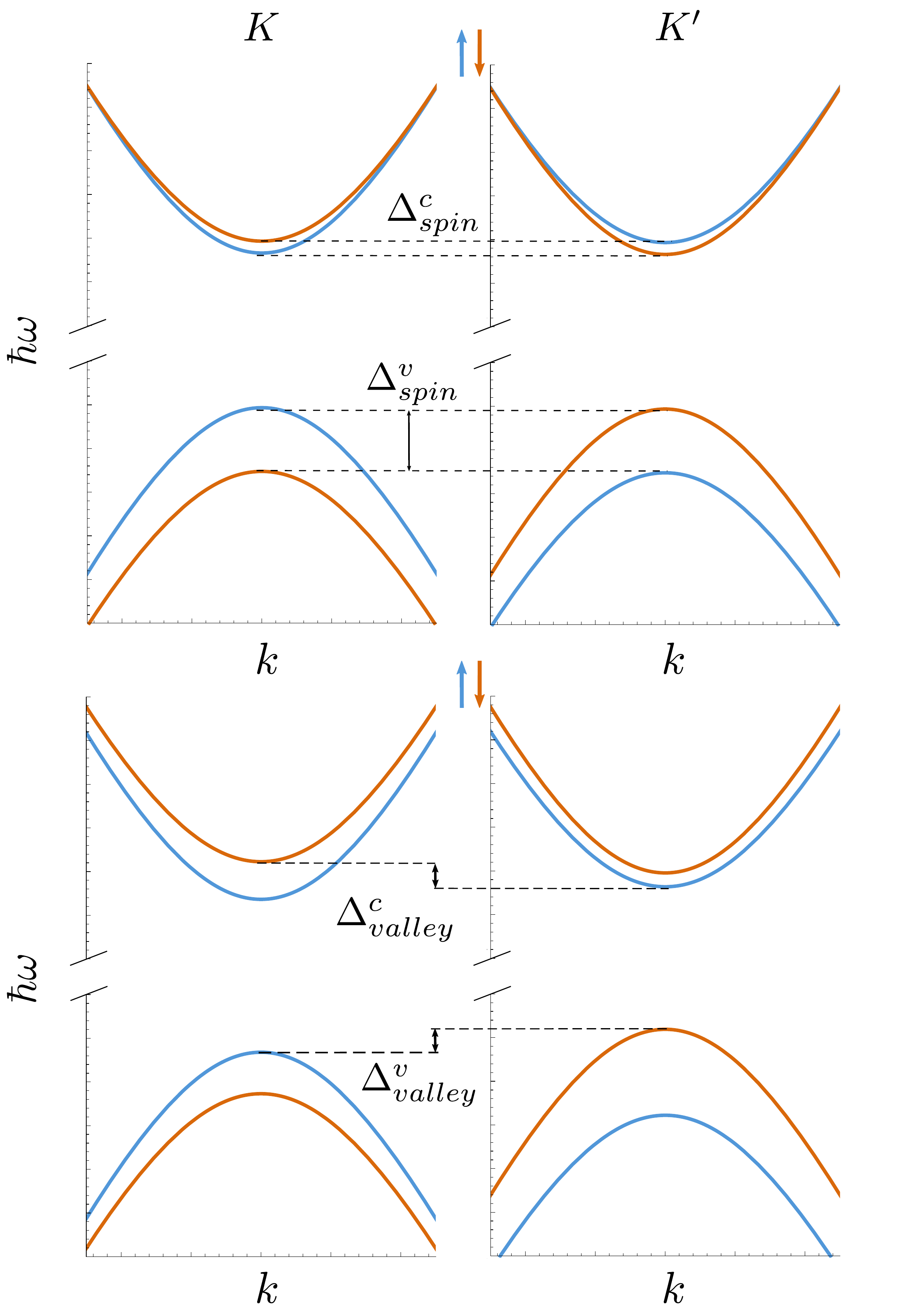}\caption{\label{fig:MoTe2 on EuO bands}Band structure near the Dirac cones
			for MoS\protect\textsubscript{2} near a Cobalt thin film, using Eq. 
			(\ref{eq:Energy Spectrum}), and the parameters of Table \ref{tab:MoTe2 parameters}. To allow an easier visualization of the different gaps the parameters of Table \ref{tab:MoTe2 parameters} were changed to $15 B_c$, $5 B_v$ and $10 \lambda_c$. These changes make the gaps more perceptible, while keeping their relative order, as well as the relative magnitude between the parameters the same.
			The figure is structured as follows: the left plots refer to the $K$
			valley, while the right ones to the $K'$ valley; the top row was
			plotted considering only spin orbit coupling effects, while the bottom
			row was plotted with both spin orbit coupling and exchange splitting
			considered. All plots are presented with the same scale, with the energies $\hbar\omega$ and wave vector $k$ given in arbitrary units. Looking at
			the top row we can see that the presence of spin orbit coupling lifts
			the spin degeneracy, and splits each band into two. The bands of spin
			up and spin down states are split by $\Delta_{spin}^{c/v}=2\lambda_{c/v}$
			for both valleys. The only difference between the two valleys is the
			swap of the spin up and spin down bands. Studying the bottom row,
			we realize that the presence of exchange splitting breaks the valley
			symmetry, and the bands that were previously aligned are now shifted
			by $\Delta_{valley}^{c/v}=2B_{c/v}$.}
\end{figure}

Using the energy spectrum obtained in Eq. (\ref{eq:Energy Spectrum}),
and the realistic parameters of Table \ref{tab:MoTe2 parameters}, we plot in
Fig. \ref{fig:MoTe2 on EuO bands} the valence and conduction bands
in the vicinity of the $K$ and $K'$ valleys, for both spin up and
spin down states. The figure is composed of two distinct cases: when
only spin orbit coupling is considered (top row); and when both spin
orbit coupling and proximity induced exchange splitting are included
(bottom row). Studying the top row, we see that we no longer have
the two degenerate spin-bands characteristic of a Dirac Hamiltonian, since
the presence of spin orbit coupling breaks the symmetry between the
two different spin states. This effect lifts the spin degeneracy and
unfolds each band in two. The energy difference between spin up and
spin down states is $\Delta_{spin}^{c}=2\lambda_{c}$ for the conduction
band, and $\Delta_{spin}^{v}=2\lambda_{v}$ for the valence band.
Comparing the bands of the $K$ and $K'$ valleys, we observe that,
although the bands associated with spin up and spin down are switched,
the valleys have a symmetric band structure, that is, only the role of the spins is interchanged between the two valleys . This is a consequence of time reversal symmetry. Studying the bottom row,
we realize that bands that were once aligned are now shifted relative
to each other due to the exchange interaction induced by magnetic proximity effect. The
relative shifts are $\Delta_{valley}^{c}=2B_{c}$, for the conduction
band, and $\Delta_{valley}^{v}=2B_{v}$, for the valence band. The
presence of this proximity effect is thus responsible for breaking
the valley symmetry, which leads to a quite different interaction
of the TMD with the two types of circular polarized light. Consequently,
this disparity in the interaction with both kinds of circular polarized
light allows us to exploit some material properties that were otherwise
inaccessible, such as a finite optical  Hall-conductivity, in the absence of a magnetic field.

\begin{center}
	\begin{table}
		\begin{centering}
			\begin{tabular}{llrll}
				\toprule 
				Variable & Value &  & Variable & Value\tabularnewline
				\cmidrule{1-2} \cmidrule{2-2} \cmidrule{4-5} \cmidrule{5-5} 
				$m$ & $1.759$ eV &  & $B_{c}$ & 1.964 meV\tabularnewline
				$\lambda_{c}$ & -1.361 meV &  & $B_{v}$ & 6.365 meV\tabularnewline
				$\lambda_{v}$ & 72.96 meV &  & $v_{F}\hbar$ & 3.52 eV$\cdot$\AA \tabularnewline
				\bottomrule
			\end{tabular}\caption{\label{tab:MoTe2 parameters} Parameters used throughout the text
				for the Hamiltonian of MoS\protect\textsubscript{2} near a Cobalt thin film. Besides the parameters shown we also considered an effective
				mass $m^{*}=0.5m_{0}$, with $m_{0}$ the bare electron mass, and
				a mean dielectric constant of the substrate (quartz) and capping layer
				(vacuum) of $\kappa=2.45$. Although these last parameters do not
				appear in the independent-particle Hamiltonian, they are necessary to compute the binding
				energies of the excitons in the considered apparatus, as they appear in the interaction potential energy. All the parameters,
				except the effective mass, were taken from Ref. [\onlinecite{Zollner2019}], where a giant exchange for MoS$_2$ on Cobalt (3 layers) was found.
				The value of the effective mass used was taken from Ref. [\onlinecite{Kormanyos2015k}]
				for MoS\protect\textsubscript{2}.}
			\par\end{centering}
	\end{table}
	\par\end{center}

We next present in Table \ref{tab:Gap Values} the gaps associated
with every considered transition, in the case where both spin orbit
coupling and exchange splitting are considered (introduced before as
$\zeta_{\tau,s_{z}}$). These values will be useful for future reference.
Analyzing the presented data, one quickly realizes that, as expected,
in the absence of exchange splitting, that is, when $B_{c}=B_{v}=0$,
the two valleys are again equivalent in terms of gap values. Furthermore,
it becomes clear that it is the difference $(B_{c}-B_{v})$ that controls actually 
the valley asymmetry.
\begin{table}
	\centering{}%
	\begin{tabular}{lll}
		\toprule 
		$(\tau,s_{z},v)\rightarrow(\tau,s_{z},c)$ & Expression $\zeta_{\tau,s_{z}}$ & Value (eV)\tabularnewline
		\midrule
		\midrule 
		$\tau=1$, $s_{z}=1$ & $m+\lambda_{c}-\lambda_{v}-B_{c}+B_{v}$ & 1.689\tabularnewline
		\midrule 
		$\tau=1$, $s_{z}=-1$ & $m-\lambda_{c}+\lambda_{v}+B_{c}-B_{v}$ & 1.829\tabularnewline
		\midrule 
		$\tau=-1$, $s_{z}=1$ & $m-\lambda_{c}+\lambda_{v}-B_{c}+B_{v}$ & 1.838\tabularnewline
		\midrule 
		$\tau=-1,s_{z}=-1$ & $m+\lambda_{c}-\lambda_{v}+B_{c}-B_{v}$ & 1.680\tabularnewline
		\bottomrule
	\end{tabular}\caption{\label{tab:Gap Values}Analytical expressions and numerical value
		for the gaps of the transitions $(\tau,s_{z},v)\rightarrow(\tau,s_{z},c)$,
		that is, the transition between the valence and conduction band, for
		a state of spin projection $s_{z}$ in the valley $\tau$ for MoS\protect\textsubscript{2}. Working
		in the limit of vertical transitions we ignore transitions from the
		valley $\tau$ to $-\tau$. No spin flips are considered. The numerical
		values were obtained using Table \ref{tab:MoTe2 parameters}. This
		table emphasizes that the asymmetry between the two valleys (seen
		on the bottom row of Fig. \ref{fig:MoTe2 on EuO bands}) is governed
		by the difference between $B_{c}$ and $B_{v}$.}
\end{table}

\section{Optical Conductivity}
\label{sec:Optical-Conductivity}

In this section, we obtain the absorbed power, from Fermi golden rule,
for both linear and circular polarized light. Afterwards, we establish
the relation between the absorbed power and the conductivity tensor. The
ultimate goal is the calculation of the optical conductivity in both the
non-interacting (see supplemental information \cite{SI}) and interacting limits for both linear and circular polarized light.

For  clarity sake, a comment on notation is now in order. When
referring to circular polarized light we will label the two orthogonal
components as $\tilde{\sigma}_{\pm}$. Later in the work we will define
conductivities also labeled by the letter $\sigma$, namely $\sigma_{\pm}$.
To avoid confusion between conductivity and polarization, it should
be clear that when a tilde ($\sim$) is used we are labeling a polarization
component, and when it is not we are referring to conductivities.

\subsection{Fermi golden rule and absorbed power}
\label{subsec:Fermi-Golden-Rule}

Let us start defining the light's electric field as

\begin{equation}
\mathbf{E}=\frac{1}{2}\left(\mathbf{E}_{0}e^{i\omega t}+\mathbf{E}_{0}^{*}e^{-i\omega t}\right),\label{eq:ElectricField Def}
\end{equation}
and work in the dipole approximation, such that the interaction between
light and matter is given by
$
H_{int}=-e\mathbf{r\cdot\mathbf{E}}.
$

From Fermi golden rule one obtains the transition rate between two different
states. If we multiply the transition rate by the energy associated
with the said transition, and sum over all initial and final states
we obtain the total power absorbed by our system, which is given by
\begin{align}
P & =\frac{2\pi}{\hbar}\frac{1}{4}\sum_{i,f}|\hbar\omega_{if}||\langle f|e\mathbf{r}\cdot\mathbf{E}_{0}|i\rangle|^{2}\times\nonumber \\
& \left[\delta(\hbar\omega-\hbar\omega_{if})+\delta(\hbar\omega+\hbar\omega_{if})\right],\label{eq:Power Absorbed}
\end{align}
where $P$ is the power absorbed; the sum is made over all initial
($i$) and final ($f)$ states; and $\hbar\omega_{if}=E_{i}-E_{f}$
is the difference between the initial and final state energies. Equation
(\ref{eq:Power Absorbed}) is valid for systems with mean occupation
numbers 1 and 0 for the initial and final states respectively, such
as, for example, the transitions between a full valence band and an
empty conduction band. If this is not the case one must add a term
which takes into account the mean occupation of the states. Although
we present two delta functions in all equations, we will only work
with positive frequencies, making one of the delta functions redundant.

Let us now consider the case of linear polarized light. The electric
field's amplitude for this type of polarization is $\mathbf{E}_{0}=E_{0}\hat{u}_{x}$, with $\hat{u}_x$ the unit vector of the $x$ axis.
Thus, the absorbed power is given by
\begin{align}
P_{x} & =\frac{\pi}{2\hbar}e^{2}E_{0}^{2}\sum_{i,f}|\hbar\omega_{if}||\langle f|x|i\rangle|^{2}\times\nonumber \\
&\left[\delta(\hbar\omega-\hbar\omega_{if})+\delta(\hbar\omega+\hbar\omega_{if})\right].\label{eq:PowerLinear}
\end{align}

For circular polarized light the amplitude of the electric field is
$\mathbf{E}_{0}=E_{0}(\hat{u}_{x}\pm i\hat{u}_{y})/\sqrt{2}$, for
$\tilde{\sigma}_{\pm}$ polarization, and the absorbed power is 
\begin{align}
P_{\pm} & =\frac{\pi}{4\hbar}e^{2}E_{0}^{2}\sum_{i,f}|\hbar\omega_{if}||\langle f|x\pm iy|i\rangle|^{2}\times\nonumber \\
& \left[\delta(\hbar\omega-\hbar\omega_{if})+\delta(\hbar\omega+\hbar\omega_{if})\right],\label{eq:PowerCirc}
\end{align}
and, with $\hat{u}_y$ the unit vector of the $y$ axis.

\subsection{Relation between the absorbed power and the conductivity}
\label{subsec:The-Non-Interacting-Case}

Now that the expressions for the power  are determined, we want to establish
their relation with the conductivity. To this end we define the current
density vector as
\begin{equation}
\mathbf{J}=\frac{1}{2}\left(\mathbf{J}_{0}e^{i\omega t}+\mathbf{J}_{0}^{*}e^{-i\omega t}\right),\label{eq:CurrentDens Def}
\end{equation}
with $\mathbf{J}_{0}=\bar{\bar{\sigma}}\mathbf{E}_{0}$, where $\bar{\bar{\sigma}}$
is the conductivity tensor given by (for an isotropic system)
\begin{equation}
\bar{\bar{\sigma}}=\left(\begin{array}{cc}
\sigma_{xx1}+i\sigma_{xx2} & \sigma_{xy1}+i\sigma_{xy2}\\
-\sigma_{xy1}-i\sigma_{xy2} & \sigma_{xx1}+i\sigma_{xx2}
\end{array}\right),
\end{equation}
where $\sigma_{xx}$ is the longitudinal conductivity, $\sigma_{xy}$
is the Hall conductivity, and the indexes 1 and 2 refer to the real
and imaginary part, respectively. We can now take advantage of these
two expressions to calculate the power absorbed in a different way.
To this end, we need to integrate over the area $A$ of our material
the dot product $\mathbf{\mathbf{J}\cdot\mathbf{E}}$, and take the
average over one period, $T$. A simplified way of writing this is
\begin{equation}
P=\langle P(t)\rangle_{T}=\frac{1}{2}\int_{A}dA\Re\left(\mathbf{\mathbf{J}_{0}^{*}\cdot\mathbf{E}_{0}}\right),
\end{equation}
with $\mathbf{E}_{0}$ and $\mathbf{J}_{0}$ defined in agreement
with Eqs. (\ref{eq:ElectricField Def}) and (\ref{eq:CurrentDens Def}).
One can now use this expression to establish the relation between
the absorbed power and different elements of the conductivity tensor
$\bar{\bar{\sigma}}.$

For linear polarized light, with $\mathbf{E}_{0}=E_{0}\hat{u}_{x}$,
one can easily obtain 
\begin{equation}
\sigma_{xx1}=\frac{2P_{x}}{AE_{0}^{2}}\label{eq:LongCond Lin}
\end{equation}
which gives a direct relation between the power absorbed with linear
polarized light and the real part of the longitudinal conductivity.

For circular polarized light, with $\mathbf{E}_{0}=E_{0}(\hat{u}_{x}\pm i\hat{u}_{y})/\sqrt{2}$,
one obtains 
\begin{equation}
\sigma_{xx1}=\frac{P_{+}+P_{-}}{AE_{0}^{2}},\label{eq:LinCond Circ}
\end{equation}
\begin{equation}
\sigma_{xy2}=\frac{P_{-}-P_{+}}{AE_{0}^{2}}.\label{eq:HallCond Circ}
\end{equation}

Now that everything is set up, we can start the explicitly calculation
the optical conductivity. For the non-interacting limit this is done in the supplemental information \cite{SI}.

\section{Excitonic effects in the Kerr angle}
\label{sec:Excitonic-Effects-in}
Until this point we have 
considered a general approach.
 From now on we will build on what has already been done
 for the non-interacting case (see supplemental information \cite{SI})
  and
expand it to the interacting case, where excitonic effects will be
considered (see the Appendix for further information on the solution of the excitonic problem). In this section, we use  the formalism developed in the Appendix for
solving the 2D Wannier equation, where we show how it can be obtained from
the Bethe-Salpeter equation (see also Ref. [\onlinecite{henriques2019absorption}]). Afterwards, we will use a semi-analytical
method to compute the longitudinal and Hall optical conductivities considering
excitonic effects, for both linear and circular polarized light. Finally, we study the effect
of excitons on the Kerr rotation angle and show that it is larger than that for thin Cobalt films by about one order of magnitude.

%
%
%
%
%\subsection{\label{subsec:Hall-Optical-Conductivity}Hall %optical conductivity
%and Kerr angle}
%
%
%
%
Continuing the work presented in Section \ref{sec:Optical-Conductivity},
 we write the conductivities\cite{henriques2019absorption}
$\sigma_{xx},$ $\sigma_{+}$, and $\sigma_{-}$ in the general form
\begin{align}
\frac{\sigma_{\mu}^{\tau,s_{z}}}{\sigma_{0}} & =4i\sum_{\nu}|\zeta_{\tau,s_{z}}+E_{\nu}|\Lambda_{\nu}^{\mu}\Bigg(\frac{1}{\hbar\omega-\zeta_{\tau,s_{z}}-E_{\nu}+i\eta}\nonumber \\
& +\frac{1}{\hbar\omega+\zeta_{\tau,s_{z}}+E_{\nu}+i\eta}\Bigg),\label{eq: Cond With Exc}
\end{align}
with $\mu=\{xx;+;-\}$; $\zeta_{\tau,s_{z}}=m+s_{z}(-B_{c}+B_{v}+\lambda_{c}\tau-\lambda_{v}\tau)$ is the gap between the
valence and conduction band for a given combination of valley, $\tau$,
and spin, $s_{z}$; $E_{\nu}$ is the exciton energy level associated
with the quantum number $\nu$ (including both the principal and magnetic
quantum numbers). Only three magnetic quantum numbers produce a non-zero
result, $m=0$ and $m=\pm2$, the largest contribution being, by far,
that of $m=0$; $\eta$ is the nonradiative decay rate, encompassing
all possible decay channels; finally, the element that differs depending
on the desired conductivity is $\Lambda_{\nu}^{\mu}$ defined as $A|\langle\nu,Q|\mathbf{r\cdot\mathbf{\hat{e}}_{\mu}}|GS\rangle|^{2}$,
with $A$ the area of the 2D material, and $\mathbf{r\cdot\mathbf{\hat{e}}_{\mu}}$
equal to $x$, $(x+iy)/\sqrt{2}$, and $(x-iy)/\sqrt{2}$, for the
cases of $\sigma_{xx},$ $\sigma_{+}$ and $\sigma_{-}$, respectively. The ket $|\nu,Q\rangle$ is defined as
\begin{align}
	|\nu,Q\rangle=\frac{1}{\sqrt{A}}\sum_\mathbf{k}\phi_\nu(\mathbf{k}) a^\dagger_{\mathbf{k}+\mathbf{Q},c} a_{\mathbf{k},v}^{}|GS\rangle
\end{align}
where $|GS\rangle$ stands for the electronic ground state of the TMD, that is, a filled valence band and an empty conduction band (for more details see the Appendix). The matrix element is promptly computed  writing the position operator
in second quantization.

\begin{figure}
\includegraphics[width=8.5cm]{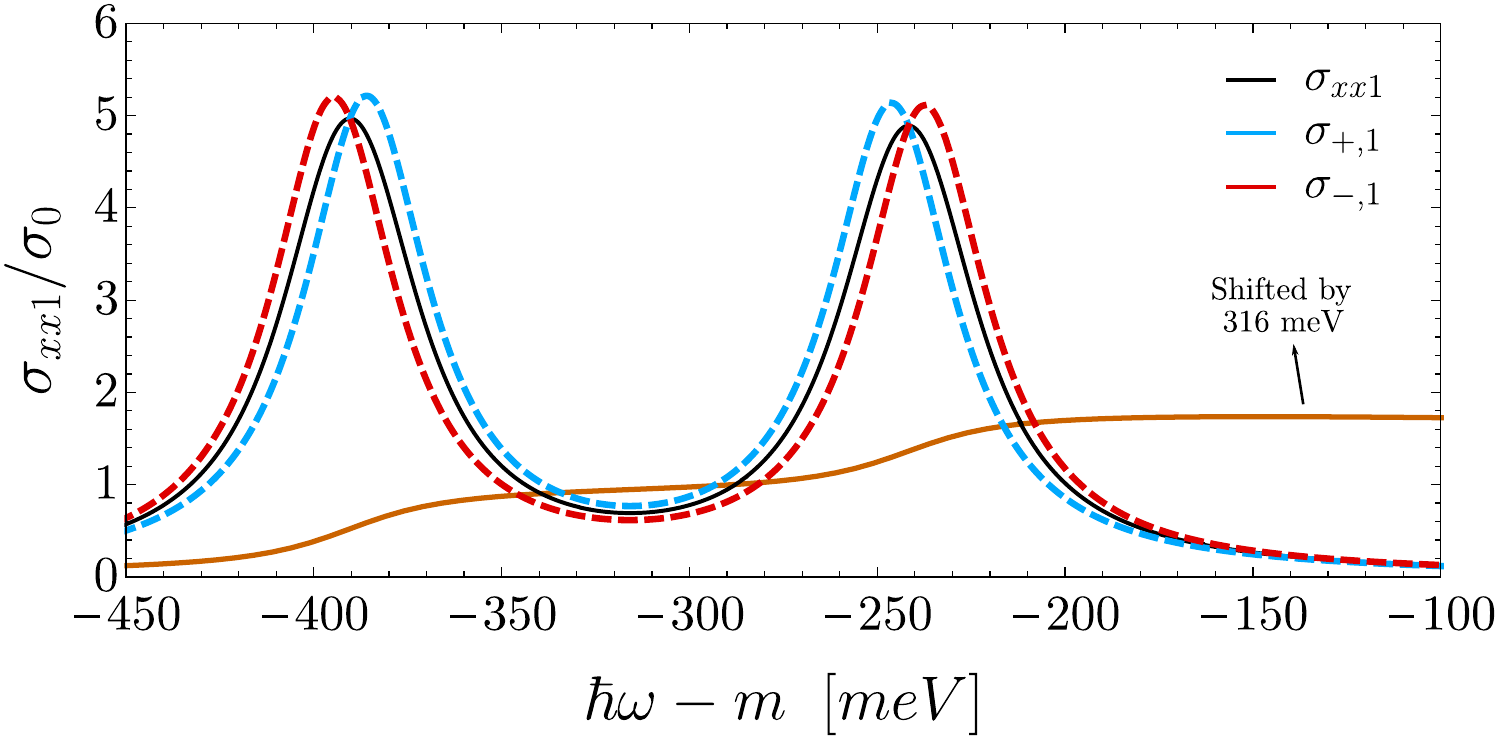}\caption{\label{fig:Cond With Exc} Representation of $\sigma_{xx1},$ $\sigma_{+,1}$
		and $\sigma_{-,1}$, defined as in Eq. (\ref{eq: Cond With Exc}).
		It is easy to see that, as expected, $\sigma_{xx1}$ can be obtained
		from the mean of $\sigma_{+,1}$ and $\sigma_{-,1}$.  The linear conductivity in the non interacting limit is also shown. It should be noted that the non interacting line was shifted by 316 meV (the exciton's binding energy) in order to bring all the plots to the same spectral region. Comparing them, one realizes that their lineshape is severely modified by the presence of excitons. It's also worthy to note that the large
		splitting ($\sim140$ meV) between the two sets of peaks is due to
		spin orbit coupling effects, while the small splitting ($\sim9$
		meV) is a consequence of the valley asymmetry induced by proximity. One final remark should
		be done: while constructing the plots for $\sigma_{+,1}$ and $\sigma_{-,1}$,
		it was observed the absence of interaction of $\sigma_{+,1}$( $\sigma_{-,1}$)
		with the $K'$ $(K)$ valley. Only excitonic peaks with quantum numbers
		$n=1$, and $m=0$ were considered. All conductivities are presented
		in units of graphene universal conductivity $\sigma_{0}=e^{2}/4\hbar$. The
		variables of Table \ref{tab:MoTe2 parameters} were used, and a nonradiative decay rate $\eta=20$ meV was considered. The precise energies
		where the excitonic resonances appear are presented in Table \ref{tab:Peaks Energy}. 
		Similar results have been reported in Ref. [\onlinecite{scharf2017magnetic}].}
\end{figure}

To compute the exciton binding energy of MoS\textsubscript{2} on
a substrate of quartz in the vicinity of a thin Cobalt film we used a mean dielectric constant $\kappa=2.45$
and a screening parameter $r_{0}=41.4$ \AA  (also for the numerical solution of the Wannier equation we have used $\Omega_{min}=-1$, $\Omega_{max}=5$,
$N=100$ and $A=6$; see see Appendix for the method of solving the Wannier equation). A binding energy of   $0.316$ eV was found for the exciton. Inserting
the computed binding energy into Eq. (\ref{eq: Cond With Exc}),
and taking the real part only, one obtains the plots presented in
Fig. \ref{fig:Cond With Exc}.

These plots show an interesting behavior, as the presence
of excitons produces major changes in the conductivities' line shape relatively to the non-interacting limit. Indeed, we can also divide
the figure in two sets of peaks separated by approximately $140$ meV. This large splitting between the
two sets of peaks is a direct consequence of spin orbit coupling,
while the small splitting within each set is due to the breaking of
valley symmetry induced by proximity.  Therefore, an experiment measuring the absorption of  circularly polarized electromagnetic radiation will be able to unveil the value of the exchange interaction. 
Using the information of Table \ref{tab:Peaks Energy},
one realizes that, when only excitons with $n=1$, and $m=0$ are
considered, the conductivities $\sigma_{+,1}$ and $\sigma_{-,1}$
are valley selective, since $\sigma_{+,1}$ only couples with the
$K$ valley ($\tau=1$), and $\sigma_{-,1}$ couples with the $K'$
valley ($\tau=-1$).

\begin{table}
	\centering{}%
	\begin{tabular}{lccc}
		\toprule 
		& $\tau$ & $s_{z}$ & $E_{g}^{\tau,s_{z}}+E_{\nu}-m$\tabularnewline
		\midrule 
		\multirow{2}{*}{$\sigma_{+}$} & 1 & 1 & $-386$ meV\tabularnewline
		& 1 & -1 & $-246$ meV\tabularnewline
		\midrule 
		\multirow{2}{*}{$\sigma_{-}$} & -1 & 1 & $-237$ meV\tabularnewline
		& -1 & -1 & $-395$ meV\tabularnewline
		\bottomrule
	\end{tabular}\caption{\label{tab:Peaks Energy}  Energies at which the excitonic resonances
appear in Fig. \ref{fig:Cond With Exc}. The data presented emphasizes
how $\sigma_{+,1}$ and $\sigma_{-,1}$ are valley selective, since the excitonic resonances of $\sigma_+$ are associated with the $K$ valley ($\tau=1$), and the ones of $\sigma_-$ are associated with $K'$ ($\tau=-1$).The origin
of the different splittings of the excitonic peaks is also made clear.}
\end{table}

\begin{figure}
\includegraphics[width=9cm]{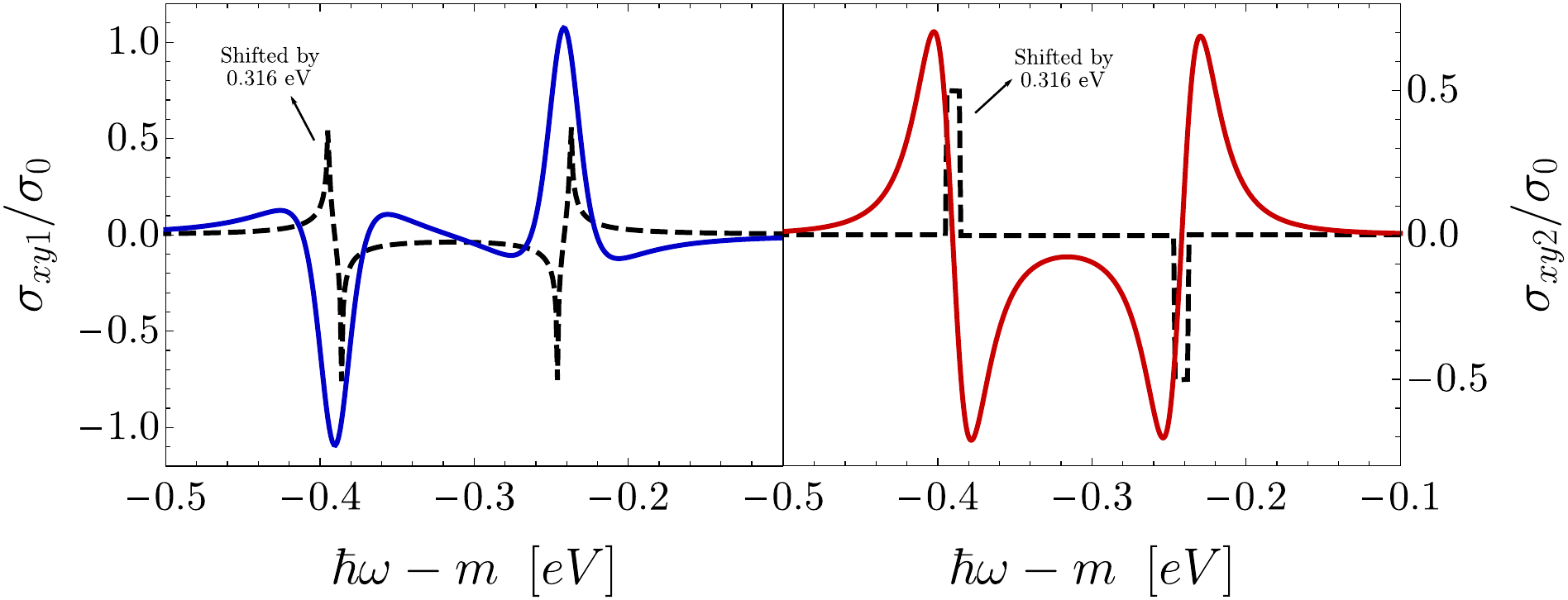}\caption{\label{fig:Hall Cond Excitons}Representation of both real (left panel) and
		imaginary (right panel) parts of Hall conductivity, obtained from Eq. (\ref{eq:Hall Cond Full Expr}).
		When we compare these plots with the ones for the non-interacting theory (represented by the dashed lines),
		we realize, quite unexpectedly, that the presence of excitonic effects
		induce a swap of behavior between the real and imaginary parts of
		Hall conductivity, while keeping a similar magnitude.  We should note that, similarly to what was done in Figure \ref{fig:Cond With Exc}, the plots of the non interacting limit were shifted by 0.316 eV, in order to show all the plots in the same energy range. All the conductivities
		are presented in units of graphene universal conductivity $\sigma_{0}=e^{2}/4\hbar.$
		The variables of Table \ref{tab:MoTe2 parameters} were used, and
		a non-radiative decay rate $\eta=20$ meV was considered  for the interacting (excitonic) calculation.
	}
\end{figure}

As was mentioned before, and is visible in Fig. \ref{fig:Cond With Exc},
the mean of $\sigma_{+,1}$ and $\sigma_{-,1}$ gives us the real
part of the longitudinal conductivity. In a similar procedure, and
according to Eq. (\ref{eq:HallCond Circ}), their difference gives
access to the imaginary part of Hall conductivity. In fact, the full
expression for $\sigma_{xy}$ is given by
\begin{equation}
\sigma_{xy}=\bigg(\frac{\sigma_{+,2}-\sigma_{-,2}}{2}\bigg)+i\bigg(\frac{\sigma_{-,1}-\sigma_{+,1}}{2}\bigg),\label{eq:Hall Cond Full Expr}
\end{equation}
where $\sigma_{\pm,1}=2P_{\pm}/(AE_{0}^{2})$, and
 the imaginary part is the expression presented in Eq. (\ref{eq:HallCond Circ}),
and the real part is readily obtained from a Kramers-Kronig transformation.
Both real and imaginary part of Hall conductivity are plotted in Fig.
\ref{fig:Hall Cond Excitons}.  In the same figure, the independent particle approximation is also depicted as thin dashed lines.

The analysis of the data depicted in the figures shows that, curiously enough, when passing
from the non-interacting case
to the one where excitonic effects are
taken into account, an (approximated) inversion of behavior occurs between the real
and imaginary part of $\sigma_{xy}$. If we compare $\sigma_{xy1}$
with excitons, with $\sigma_{xy2}$ in the non-interacting case, the
former appears to be a smoothed out version of the latter. The same
goes for $\sigma_{xy2}$ with excitons, and $\sigma_{xy1}$ without
them. 

The difference between the optical conductivities computed with and without excitonic effects can be understood taking a closer look at Fig.  \ref{fig:Cond With Exc}. Looking at
the difference between $\sigma_{-,1}$ and $\sigma_{+,1}$ we see
that, at lower energies, both conductivities are close to zero. Then,
$\sigma_{-,1}$ starts to outgrow $\sigma_{+,1}$ until the former
reaches a maximum, which means that $\sigma_{xy2}$ also reaches a maximum. Next, $\sigma_{-,1}$ decreases, and $\sigma_{+,1}$
increases, which leads to $\sigma_{xy2}=0$ when the lines intercept,
and afterwards to a minimum of $\sigma_{xy2}$, when $\sigma_{+,1}$ reaches its maximum. Subsequently, $\sigma_{-,1}$ decreases,
and both conductivities are, again, close to zero. From this point
onward the process happens in reverse order. Comparing now this description
with Fig. \ref{fig:Hall Cond Excitons} we realize that this is
precisely the behavior presented by $\sigma_{xy2}$.

\begin{figure}
		\includegraphics[width=9cm]{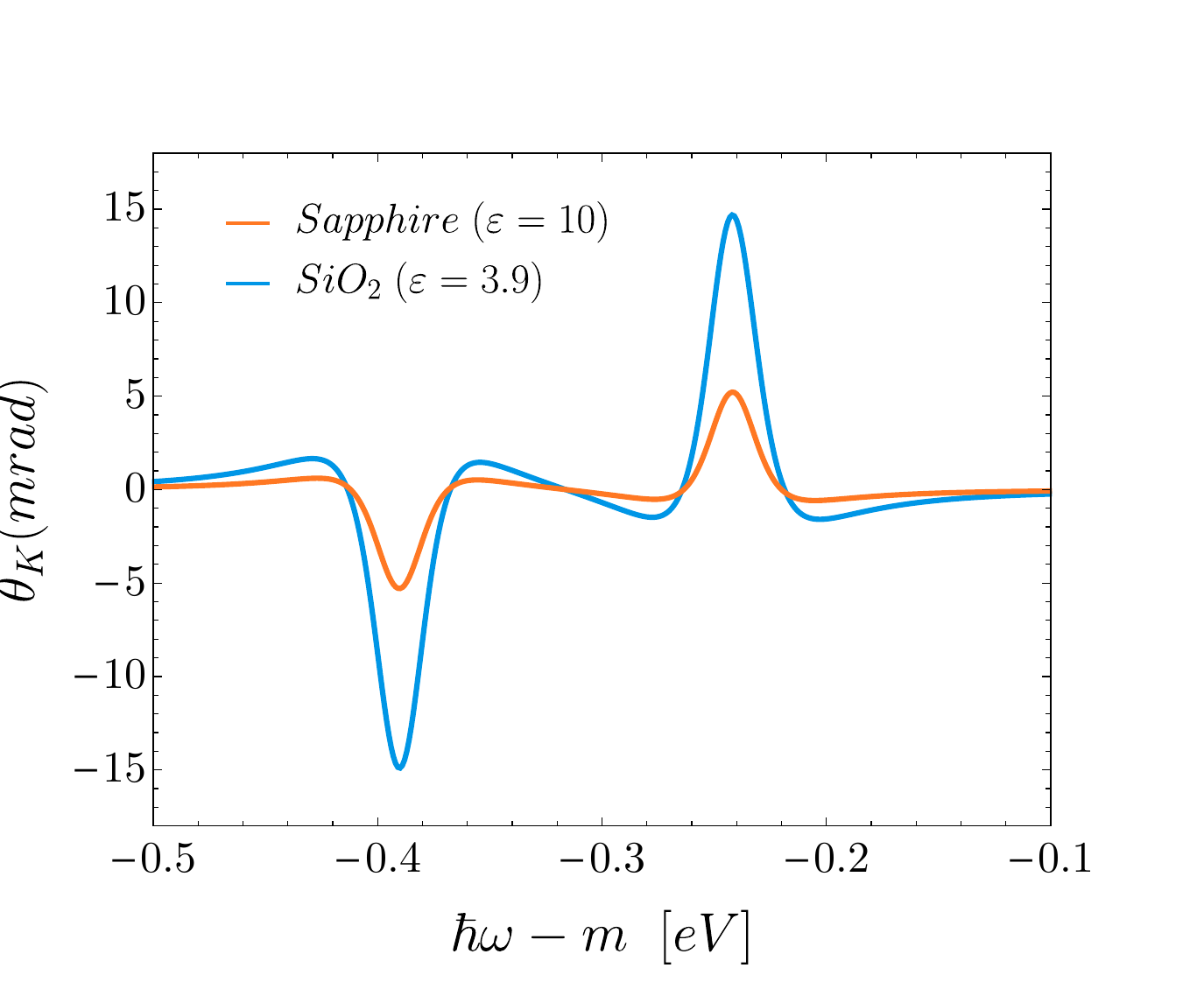}\caption{\label{fig:Kerr Angle Plot}
		 Representation of Kerr angle, when
			excitonic effects are considered, using Eq. (\ref{fig:Kerr Angle Plot}),
			and the conductivities previously obtained in Fig. \ref{fig:Hall Cond Excitons}. We show the Kerr angle for two substrates, quartz and sapphire, and verify that as the substrate dielectric constant increases, the Kerr angle intensity decreases. Although in principle one should consider the effects of the hBN layer and cobalt thin film, we consider the substrate contribution as the dominant one, and discard the effect of the others two materials.}
\end{figure}

Now that the conductivities are determined, we can move on to the
calculation the effect of excitons on the Kerr angle. It can be shown
(Ref. [\onlinecite{GoncaloCatarina2019}]) that, in the limit of small angles, the Kerr angle is related
to the linear and Hall conductivity through the following equation
\begin{equation}
\theta_K = \Re \Bigg( \frac{2c\mu_0 \sigma_{xy}}{(\varepsilon-1)+\Sigma} \Bigg), \label{eq:Kerr Angle}
\end{equation}
where $\varepsilon$ is the dielectric constant of the substrate,
$c$ is the light speed in vacuum, $\mu_{0}$ is the vacuum permittivity,
\begin{equation}
\Sigma=2\sqrt{\varepsilon}c\mu_0 \sigma_{xx} + c^2\mu_0^2(\sigma_{xx}^2+\sigma_{xy}^2) + 2ic\mu_0\sigma_{xy}\,,
\end{equation}
and $\sigma_{xx}$ and $\sigma_{xy}$ are the conductivities previously defined. We should note that in Ref.\cite{raja2017coulomb} it has been experimentally shown that even the presence of a single graphene layer is enough to substantially change the exciton binding energies. Here, due to the complexity of the considered apparatus, it is no easy task to give a full description of the effects of the hBN layer, the cobalt thin film and the substrate on the TMD. We do believe, however, that the effect of the substrate dominates over all other, and thus only consider its contribution to the problem. Using Eq. (\ref{eq:Kerr Angle}) and the conductivities formerly
obtained, we compute the Kerr angle plotted in Fig. \ref{fig:Kerr Angle Plot}.

As expected, since $\varepsilon$ is far greater than any other element
on the denominator of Eq. (\ref{eq:Kerr Angle}), the Kerr angle
takes its shape from the real part of Hall conductivity.  We can also see that as the environment's dielectric screening increases (when passing to a sapphire substrate) the Kerr angle intensity decreases. In what magnitude
is concerned, we obtain a Kerr angle orders of magnitude higher then
the ones from Refs. \cite{perov2010terahertz,yang2006spectral}, where a 2D electron gas is studied (at 10 K in the case of Ref. \cite{yang2006spectral}). If the broadening parameter was reduced then the Kerr angle magnitude would be even larger. The $20$ meV considered in this work is a conservative value, and TMD's encapsulated in hBN can have broadenings as low as 2 meV \cite{cadiz2017excitonic}.

We also note that in Ref. \cite{Hamidi2014} the Kerr angle for a cobalt film $4$ nm thick was measured presenting a magnitude one order of magnitude smaller when compared to the values of Figure \ref{fig:Kerr Angle Plot}. Therefore, an experiment made in the system proposed in this work will essentially probe Kerr effect due to the TMD.

In Ref. [\onlinecite{lee2016electrical}] the Kerr angles was studied for
bilayer MoS\textsubscript{2}. In this work an electric field was
applied perpendicularly to the samples, which led to the spatial separation
of the $K$ and $K'$ valleys. When a spatial map of the Kerr angle
was constructed it was observed negative values for the Kerr angle,
followed by positive ones, similarly to the plot of Fig. \ref{fig:Kerr Angle Plot}.

\section{Conclusions}

In this paper we have studied the effects produced on MoS$_2$ band structure due to
proximity to a Cobalt ferromagnetic thin film (3 layers), more specifically, the case of
MoS\textsubscript{2} on a heterostructure composed of MoS\textsubscript{2}/hBN/Co/quartz was studied.

We started  using an effective low energy Hamiltonian, composed
of a massive Dirac Hamiltonian, a spin orbit coupling term, and an
exchange contribution, to theoretically describe the changes that
the TMD band structure undergoes when placed in the vicinity of a
ferromagnetic thin film. We have verified that, in the presence of magnetic
proximity effects, the valleys $K$ and $K'$, have an asymmetric
band structure, leading to different interaction with the two components
of circular polarized light.

Using Fermi's golden rule, we then proceeded to the computation of
the longitudinal conductivity $\sigma_{xx}$, and Hall conductivity
$\sigma_{xy}$ in the non-interacting limit (the results can be found in the supplemental information \cite{SI}). The existence of a finite
Hall conductivity is a direct consequence of the asymmetry between
the $K$ and $K'$ valleys produced by proximity effects. Afterwards,
we presented a semi-analytical method that allowed us to extend
our work in the non-interacting limit to the case where excitons
are present. Using this method we, once more, computed the optical conductivities, but now taking into consideration excitonic
effects. Once again, we obtained a non-zero Hall conductivity due
to the different interaction of the two components of circular polarized
light with the TMD. An intriguing finding appeared when the plots
of the Hall conductivity in the non-interacting and in the interacting limits
were compared. Looking at Fig. \ref{fig:Hall Cond Excitons},
it is possible to see that an inversion of behavior between the real
and imaginary part of the Hall conductivity takes place when passing
between these two limits, that is, if we compare $\sigma_{xy1}$ with
excitons, with $\sigma_{xy2}$ in the non-interacting case, the former
appears to be a smoothed out version of the latter. The same happens
for $\sigma_{xy2}$ with excitons and $\sigma_{xy1}$ without them. This inversion is a direct consequence of the dramatic change of the line shape of the optical conductivity when excitonic effects are included.

Finally, we used the Hall conductivity containing excitonic effects
to obtain the Kerr rotation angle. Since the Kerr rotation angle has
a direct dependence on $\sigma_{xy}$, it can only be explored when
an asymmetry between the valleys is induced, that is, when the TMD is 
proximitized.

\section*{acknowledgements}
	N.M.R.P. acknowledges support from the European Commission through
	the project \textquotedblleft Graphene-Driven Revolutions in ICT and
	Beyond\textquotedblright{} (Ref. No. 785219), and the Portuguese Foundation
	for Science and Technology (FCT) in the framework of the Strategic
	Financing UID/FIS/04650/2019. In addition, N. M. R. P. acknowledges
	COMPETE2020, PORTUGAL2020, FEDER and the Portuguese Foundation for
	Science and Technology (FCT) through projects PTDC/FIS- NAN/3668/2013
	and POCI- 01-0145-FEDER-028114, and POCI-01-0145-FEDER- 029265 and
	PTDC/NAN-OPT/29265/2017, and POCI-01-0145-FEDER-02888.
	G. C. acknowledges Funda\c{c}\~{a}o para a Ci\^{e}ncia e a Tecnologia (FCT) for Grant No. SFRH/BD/138806/2018.
	G. C. and J. F.-R. acknowledge financial support from FCT through Grant No. P2020-PTDC/FIS-NAN/4662/2014.
	J. F.-R. acknowledges financial support
	from FCT for the P2020-PTDC/FIS-NAN/4662/2014, the P2020-PTDC/FIS-NAN/3668/2014 and the UTAPEXPL/NTec/0046/2017 projects, as well as Generalitat
	Valenciana funding Prometeo2017/139 and MINECO Spain (Grant No. MAT2016-78625-C2).

\appendix

\section{\label{subsec:From-BSE-to}
From BSE to the Wannier equation and its solution}

In second quantization, the state of an exciton of momentum $\mathbf{Q}$,
in motion in a TMD monolayer of area $A$, can be written as
\begin{equation}
\vert\nu,\mathbf{Q}\rangle=\frac{1}{\sqrt{A}}\sum_{\mathbf{k}}\phi_{\nu}(\mathbf{k})a_{\mathbf{k}+\mathbf{Q},c}^{\dagger}a_{\mathbf{k},v}\vert GS\rangle,\label{eq:Exciton State}
\end{equation}
where the state $\vert GS\rangle$ represents the electronic ground
state of the TMD, that is, a filled valence band and an empty conduction
band; $\phi_{\nu}(\mathbf{k})$ is the Fourier transform of the real
space exciton wave function, with $\nu$ representing the principal
and magnetic quantum numbers that characterize it; the second quantized
operators $a_{\mathbf{k}+\mathbf{Q},c}^{\dagger}$ and $a_{\mathbf{k},v}$
create and annihilate an electron of momentum $\mathbf{k+\mathbf{Q}}$
in the conduction band and an electron of momentum $\mathbf{k}$ in
the valence band, respectively. This state can be expressed in a condensed
form as $\vert\nu,\mathbf{Q}\rangle=b_{\mathbf{Q},\nu}^{\dagger}\vert GS\rangle$,
with $\textbf{\ensuremath{b_{\mathbf{Q},\nu}^{\dagger}}}$ the bosonic
operator, defined as
\begin{equation}
b_{\mathbf{Q},\nu}^{\dagger}=\frac{1}{\sqrt{A}}\sum_{\mathbf{k}}\phi_{\nu}(\mathbf{k})a_{\mathbf{k}+\mathbf{Q},c}^{\dagger}a_{\mathbf{k},v}.\label{eq:Bosonic Operator}
\end{equation}
We note that the bosonic nature of the operator (\ref{eq:Bosonic Operator})
is guaranteed only in an average over the ground state.

The electrons in the TMD monolayer are described by the Hamiltonian
$
H=H_{0}+V,
$
where
\begin{equation}
H_{0}=\sum_{\lambda,\mathbf{k}}E_{\lambda,\mathbf{k}}\hat{a}_{\lambda,\mathbf{k}}^{\dagger}\hat{a}_{\lambda,\mathbf{k}},
\end{equation}
with $\lambda=\{c,v\}$, $E_{\lambda,\mathbf{k}}=E_{\alpha}$, and
with the interaction term given by
\begin{align}
V & =\frac{1}{2A}\sum_{\mathbf{k}_{1},\mathbf{k}_{2},\mathbf{p}}\sum_{\lambda_{1}\lambda_{2}\lambda_{3}\lambda_{4}}V(\mathbf{p})F_{\lambda_{1}\lambda_{2}\lambda_{3}\lambda_{4}}(\mathbf{k}_{1},\mathbf{k}_{2},\mathbf{p})\nonumber \\
 & \times\hat{a}_{\mathbf{k}_{1}+\mathbf{p},\lambda_{1}}^{\dagger}\hat{a}_{\mathbf{k}_{2}-\mathbf{p},\lambda_{2}}^{\dagger}\hat{a}_{\mathbf{k}_{2},\lambda_{3}}\hat{a}_{\mathbf{k}_{1},\lambda_{4}},
\end{align}
and
\begin{equation}
F_{\lambda_{1}\lambda_{2}\lambda_{3}\lambda_{4}}(\mathbf{k}_{1},\mathbf{k}_{2},\mathbf{p})=u_{\mathbf{k}_{1}+\mathbf{p},\lambda_{1}}^{\dagger}u_{\mathbf{k}_{2}-\mathbf{p},\lambda_{2}}^{\dagger}u_{\mathbf{k}_{2},\lambda_{3}}u_{\mathbf{k}_{1},\lambda_{4}},
\end{equation}
is a product of the spinors presented in equations (2)
and (3) from the main text and is termed the form factor. It is
important to note that we do not specify the spin and valley indexes
of the spinors, since the procedure is identical for any spin and
valley combination. The function $V(\mathbf{p})$ is the Fourier transform
of the Rytova-Keldysh potential given in equation (5),
and whose expression is known analytically \citep{rytova1967}
\begin{equation}
V(\mathbf{q})=\frac{e^{2}}{2\epsilon_{0}q(r_{0}q+\kappa)}.
\end{equation}
We now intend to show that if the state $\vert\nu,\mathbf{Q}\rangle$
is an eigenstate of $H$, then $\phi_{\nu}(\mathbf{k})$ obeys the
Bethe-Salpeter equation, which when Fourier transformed to the real
space becomes the Wannier equation, under a set of simplifying assumptions.

As previously said, with the purpose of obtaining the Bethe-Salpeter
equation, we start assuming that the state (\ref{eq:Exciton State})
is an eigenstate of the Hamiltonian $H$. If this is true, then $H$
can be written as $H=\sum_{\mathbf{Q},\nu}E_{\mathbf{Q},\nu}b_{\mathbf{Q},\nu}^{\dagger}b_{\mathbf{Q},\nu}$,
where $E_{\mathbf{Q},\nu}$ are the energy eigenvalues of the exciton.
Afterwards, we evaluate the commutator of $H$ with $b_{\mathbf{Q},\nu}^{\dagger}$
using both the fermionic (when dealing with the $a$ and $a^{\dagger}$
operators) and bosonic (when dealing directly with the $b$ and $b^{\dagger}$
operators) representations. In the end we demand that both results
must be equal. Following this procedure, one obtains the equation
for the wave function of the exciton in momentum space:
\begin{align}
E\phi_{\nu}(\mathbf{k}) & =\phi_{\nu}(\mathbf{k})(E_{c,\mathbf{k}}-E_{v,\mathbf{k}})+\frac{1}{A}\phi_{\nu}(\mathbf{k})\sum_{\mathbf{p}}V(\mathbf{p})\nonumber \\
 & \times\left[u_{\mathbf{k},v}^{\dagger}u_{\mathbf{k-p},v}^{\dagger}u_{\mathbf{k},v}u_{\mathbf{k-p},v}-u_{\mathbf{k},c}^{\dagger}u_{\mathbf{k+p},v}^{\dagger}u_{\mathbf{k},c}u_{\mathbf{k+p},v}\right]\nonumber \\
 & -\frac{1}{A}\sum_{\mathbf{p}}V(\mathbf{p})\phi_{\nu}(\mathbf{p+k})u_{\mathbf{p+k},v}^{\dagger}u_{\mathbf{k},c}^{\dagger}u_{\mathbf{p+k},c}u_{\mathbf{k},v}
\end{align}
where $E$ represents the exciton energy eigenvalues. This equation
in momentum space for $\phi_{\nu}(\mathbf{k})$ is known as the Bethe-Salpeter
equation. Analyzing it, one realizes that each term has
a clear and distinct meaning. While the first term gives the energy
of a particle-hole excitation when no interactions are treated, the
second term represents the exchange energy correction to the non-interacting
particle-hole excitation energy; its value determines the magnitude
of the gap. The third and final term describes the attraction between
the electron and hole present in the conduction and in the valence
band, respectively. Crucially this term is negative, although in original
Hamiltonian the interaction between electrons is obviously repulsive.
To solve this equation directly in the momentum space one would have
to solve an integral equation, that, although possible, can be a rather
delicate process \citep{chaves2017excitonic}. Another way of solving
this problem passes by converting this integral equation into a differential
one, going to real space by means of a Fourier transform. Unfortunately,
another complication arises, since the spinors product in the third
term of the Bethe-Salpeter equation makes Fourier transforming an
almost impossible task. In order to solve this ravel, we make the
following observation concerning the form factors. In the case of
a large energy gap, one can take

\begin{equation}
u_{\mathbf{p+k},v}^{\dagger}u_{\mathbf{k},c}^{\dagger}u_{\mathbf{p+k},c}u_{\mathbf{k},v}\longrightarrow1+\mathcal{O}\left(1/m^{2}\right)
\end{equation}
so as to forego the spinorial structure of the last term of the Bethe-Salpeter
equation. Although the reader, recalling the values of Table 1,
may find this approximation to crude, we actually found  excellent agreement
with the results from Ref. \citep{scharf2017magnetic} when the case of MoTe$_2$ on a substrate of EuO was studied. It is also
considered that both the energy difference, $E_{c,\mathbf{k}}-E_{v,\mathbf{k}}$,
and the exchange energy corrections are expanded up to second order
in $\mathbf{k}$. The resulting differential equation -{}-{}-in real
space-{}-{}- reads

\begin{equation}
(E-E_{g})\psi_{\nu}(\mathbf{r})=-\bigg[\frac{\hbar^{2}}{2\mu}\nabla^{2}+V(\mathbf{r})\bigg]\psi_{\nu}(\mathbf{r})
\end{equation}
which is known as the Wannier equation for the excitonic wave function.
In this context, $\mu$ is the reduced mass of the exciton which,
in our model, reads $m^{\ast}/2$, with $m^{*}$ the effective mass
of the electron/hole, and $E_{g}$ is the corrected gap considering
the exchange correction. In this work, however, for simplicity sake,
we will discard the exchange correction, and consider $E_{g}$ as
the gap given by Eq. (4) of the main text.
This decision allows us to better compare our results with the ones
from Ref.\citep{scharf2017magnetic}. If the exchange correction to
the gap was considered, the separation between the valence and conduction
bands would increase, the conductivities would appear at higher energies,
and their magnitude would be bigger.

In the previous paragraphs we have shown that in order to solve the
excitonic problem one needs to first obtain the excitonic wave function
in both real and reciprocal space. Here, we show that a quasi-analytical
expression for the wave functions of the exciton can be written using
a set of Gaussian functions, which simplifies the calculations when
compared to a fully numerical method. Although other possibilities
exist for the choice of basis, such as the Slater basis, the choice
of a Gaussian basis is the approach used in this this work. Inspired
by the solution of the 2D hydrogen atom \citep{yang1991analytic},
we write our wave function as:
\begin{equation}
\psi_{\nu}(\mathbf{r})=\mathcal{A}_{\nu}\sum_{j=1}^{N}c_{j}^{\nu}e^{im\theta}r^{|m|}e^{-\zeta_{j}r^{2}},
\end{equation}
where $e^{im\theta}r^{|m|}$ follows from the eigenfunctions of the
$z-$component of the angular momentum and the behavior of the radial
wave function near the origin, for $m=0,\pm1,\pm2,\ldots$, the magnetic
quantum number; the Gaussian term $e^{-\zeta_{j}r^{2}}$ describes
the decay of the wave function far from the origin, with a decay constant
dependent on $\zeta_{j}$. The coefficients $c_{j}^{\nu}$ weight
the different terms; ${\cal A}_{\nu}$ is a normalization constant
given by
\begin{equation}
\mathcal{A}_{\nu}=\sqrt{\frac{1}{\pi\:\mathcal{S}_{\nu}}},
\end{equation}
with $\mathcal{S}_{\nu}=\sum_{j=1}^{N}\sum_{j'=1}^{N}c_{j}^{\nu*}c_{j'}^{\nu}(\zeta_{j}+\zeta_{j'})^{-1-|m|}\Gamma(|m|+1)$,
and $\Gamma(x)$ the gamma-function. An additional advantage of this
method is that the matrix elements of both the kinetic operator and
the electron-electron interaction do not mix different $m$ values
and therefore, the eigenvalue problem is block diagonal in the angular
momentum space.

Using our trial wave function and computing the matrix elements of
the kinetic and potential energy operators, the generalized eigenvalue
problem acquires the form
\begin{equation}
\sum_{j=1}^{N}[H(\zeta_{i},\zeta_{j})-S(\zeta_{i},\zeta_{j})E]c_{j}^{\nu}=0,\label{eq:GHW}
\end{equation}
where $H(\zeta_{i},\zeta_{j})$ is called the Hamiltonian kernel and
$S(\zeta_{i},\zeta_{j})$ is the superposition kernel. The latter
differs from a Kronecker$-\delta$ kernel since the set of Gaussian
functions is not an orthogonal basis. Both kernels have an analytical
expression given in Ref. \citep{henriques2019absorption}. Equation
(\ref{eq:GHW}) has first been written in nuclear physics and is termed
the Griffin-Hill-Wheeler equation \citep{griffin1957collective}.
The key aspect of the method is the sensible choice of the parameters
$\zeta_{j}$. A choice not so well known is the use of a logarithmic
grid of $\zeta's$ according to the rule given in ref. \citep{mohallem1986further}
$
\Omega=\frac{\ln\zeta}{A},
$
where $A>1$ and the $\Omega's$ are uniformly spaced in an interval $[\Omega_{{\rm min}},\Omega_{{\rm max}}]$
and $A$ is typically chosen between 6 and 8. The exposed method was
previously used in ref. \citep{henriques2019absorption}, and was
shown to produce excellent results.

%\bibliographystyle{apsrev4-1}
%\bibliography{References}


\end{document}